\documentclass{jpp}

\usepackage{amsmath}
\usepackage{cite}
\usepackage{bm}
\usepackage{mathrsfs}

\newcommand{\mno}{\mathnormal{\Omega}}
\newcommand{\mng}{\mathnormal{\Gamma}}

\ifprodtf \else
  \checkfont{eurm10}
  \iffontfound
    \IfFileExists{upmath.sty}
      {\typeout{^^JFound AMS Euler Roman fonts on the system,
                   using the 'upmath' package.^^J}%
       \usepackage{upmath}}
      {\typeout{^^JFound AMS Euler Roman fonts on the system, but you
                   dont seem to have the}%
       \typeout{'upmath' package installed. JFM.cls can take advantage
                 of these fonts,^^Jif you use 'upmath' package.^^J}%
      }
  \else
  \fi
\fi

\ifprodtf \else
  \checkfont{msam10}
  \iffontfound
    \IfFileExists{amssymb.sty}
      {\typeout{^^JFound AMS Symbol fonts on the system, using the
                'amssymb' package.^^J}%
       \usepackage{amssymb}%

	   \renewcommand{\simeq}{\approx}
      }{}
  \else
  \fi
\fi

\ifprodtf \else
  \IfFileExists{amsbsy.sty}
    {\typeout{^^JFound the 'amsbsy' package on the system, using it.^^J}%
     \usepackage{amsbsy}}
    {}
\fi

\newsavebox{\astrutbox}
\sbox{\astrutbox}{\rule[-5pt]{0pt}{20pt}}

\mathchardef\varLambda="0103

\title[Wave-kinetic description of nonlinear photons]%
{Wave-kinetic description \\ of nonlinear photons} 

\author[M. Marklund, P.\,K. Shukla, G. Brodin and L. Stenflo]{M\ls
  A\ls T\ls T\ls I\ls A\ls S\ns M\ls A\ls R\ls K\ls L\ls U\ls N\ls
  D$^1$\thanks{E-mail address: marklund@elmagn.chalmers.se}\thanks{Also at:
    Department of Electromagnetics, Chalmers 
    University of Technology, SE--412 96 G\"oteborg, Sweden}, 
  P\ls A\ls D\ls M\ls A\ns K.\ns S\ls H\ls U\ls K\ls L\ls A$^1$%
  \thanks{Also at: Department of Physics, Ume{\aa} University, SE--901 87
  Ume{\aa}, Sweden}, G\ls
  E\ls R\ls T\ns B\ls R\ls O\ls D\ls I\ls N$^2$\ns \and \ns L\ls E\ls
  N\ls N\ls A\ls R\ls T\ns S\ls T\ls E\ls N\ls F\ls L\ls O$^2$}

\affiliation{$^1$ Institut f\"ur Theoretische Physik IV, Fakult\"at f\"ur
  Physik und Astronomie, Ruhr-Universit\"at Bochum, D--44780 Bochum,
  Germany\\[\affilskip]
  $^2$ Department of Physics, Ume{\aa} University, SE--901 87
  Ume{\aa}, Sweden}

\begin{document}

\maketitle

\begin{abstract}
  The nonlinear interaction, due to quantum
  electrodynamical (QED) effects, between photons is investigated
  using a wave-kinetic description. Starting from a coherent wave
  description, we use the Wigner transform technique to obtain a set
  of wave-kinetic equations, the so called Wigner--Moyal
  equations. These equations are coupled to a background radiation fluid, whose
  dynamics is determined by an acoustic wave equation. In the slowly
  varying acoustic limit,  
  we analyse the resulting system of kinetic equations, and show that they
  describe instabilities, as well as Landau-like damping. 
  The instabilities may lead to break-up and focusing
  of ultra-high intensity multi-beam systems, which in conjunction
  with the damping may result in stationary strong field
  structures. The results could be of relevance for the next
  generation of laser-plasma systems.   
\end{abstract}


\section{Introduction}

Currently, the development of laser technology and laser-plasma
accelerators is pushing the limits of the achievable field strengths
in laboratories to levels unprecedented in human history
\cite{Mourou-Barty-Perry,Pukhov,Bulanov-Esirkepov-Tajima,bob,bob2}. The
successes in laser-plasma based acceleration may even hold the promise
of reaching the critical Schwinger limit, when the vacuum becomes
fully nonlinear 
\cite{Bulanov-Esirkepov-Tajima}. Thus, it is clear that the nonlinear
quantum electrodynamical (QED) vacuum effects will 
become important. Another possibility that has been pointed out is the
formation of plasma channels, evacuated
plasma cavities which could support immense field strengths. It has been
suggested that elastic photon--photon scattering could be detected
within these system, using the next generation of laser-plasma facilities
\cite{Shen-Yu,Shen-etal}.
Moreover, a large number of astrophysical systems,
such as magnetars \cite{magnetar}, gives rise to more extreme
conditions than one could ever produce in earth-based laboratories. As
an example of nonlinear QED effects, the possibility of photon--photon
scattering is perhaps the most prominent
\cite{Heisenberg-Euler,Weisskopf,Schwinger}. There has been much
interest in this particular effect, both from an experimental and an
astrophysical point of view (see, e.g., 
\cite{Bialynicka-Birula,Adler,Harding,Ding-Kaplan1,Latorre-Pascual-Tarrach,%
  Dicus-Kao-Repko,Ding-Kaplan2,Soljacic-Segev,Brodin-etal,%
  Brodin-marklund-Stenflo,Brodin-Marklund-Stenflo2,Boillat,Heyl-Hernquist,%
  Heyl-Hernquist2,Shaviv-Heyl-Lithwick,Denisov-Svertilov,%
  DeLorenci-Klippert-Novello,Thoma} and
references therein, and \cite{Marklund-Brodin-Stenflo} for an
up-to-date discussion and \cite{Greiner-Muller-Rafaelski} for an
overview). The formulation of this effect in terms of the
Heisenberg--Euler Lagrangian has been used to investigate such diverse
topics as photon splitting and magnetic lensing
\cite{Heyl-Hernquist,Heyl-Hernquist2,Shaviv-Heyl-Lithwick,Denisov-Svertilov}, 
direct detection via second harmonic 
generation \cite{Ding-Kaplan2}, self-focusing \cite{Soljacic-Segev}, nonlinear
wave mixing in cavities
\cite{Brodin-marklund-Stenflo,Brodin-Marklund-Stenflo2} and  
waveguide propagation \cite{Brodin-etal}.
The approach presented here will be of importance for both
the experimental and theoretical questions that may be posed
concerning photon--photon scattering.

\section{The governing equations}

Let us here start by giving a short review of the necessary
equations. This will serve as a guide for the steps to follow. We
thus set up a coupled system of nonlinear Schr\"odinger equations
(NLSE) for
two electromagnetic pulses, and the corresponding acoustic wave
equation for the fluid background. We then apply the Wigner
transformation to the NLSE, and obtain a coupled set of Vlasov-like
equations, the Wigner--Moyal equations. In previous work
\cite{Shukla-Eliasson,two-pulse}, it has been found that the effect of
photon--photon scattering is to introduce modulational and
filamentational instabilities in coherent photon
systems. Here we will show that the Wigner--Moyal system also
leads to modulational instabilities, and that these can be constrained
by Landau-like damping. The results and
applications thereof are then discussed.

The nonlinear self-interaction of photons can be expressed in terms 
of the Heisenberg--Euler effective Lagrangian
\cite{Heisenberg-Euler,Weisskopf,Schwinger}
\begin{equation}\label{eq:lagrangian}
  L = \varepsilon_0F +
  \kappa\varepsilon_0^2\left[4F^2 + 7G^2 \right],
\end{equation}
where $F = (E^2 - c^2B^2)/2$ and $G = c\bm{E}\cdot\bm{B}$.
The parameter $\kappa \equiv 2\alpha^2\hbar^3/45m_e^4c^5 \approx 1.63\times
10^{-30}\, \mathrm{m}\mathrm{s}^{2}/\mathrm{kg}$ 
represents the inverse of a critical energy density.
Here $\alpha$ is
the fine-structure
constant, $\hbar$ is the Planck constant, $m_e$ the electron mass, and
$c$ the velocity of light in vacuum. The Lagrangian
(\ref{eq:lagrangian}) is valid
when there is no electron--positron pair creation and the field
strength is smaller than the critical field, i.e., 
\begin{equation}
  \omega \ll m_ec^2/\hbar \,\,\text{ and }\,\, |\bm{E}| \ll
  E_{\text{crit}} \equiv 
  m_ec^2/e\lambda_c   
  \label{eq:constraint} 
\end{equation}
respectively. Here $e$ is the elementary charge, $\lambda_c$ is the
Compton wave length, and $E_{\text{crit}} \simeq
10^{18}\,\mathrm{V}/\mathrm{m}$.  

In Ref.\ \cite{two-pulse}, the coupled equations 
\begin{subequations}
\begin{eqnarray}
  i\left( \frac{\partial}{\partial t} +
  c\hat{\bm{k}}_{01}\cdot\nabla
  \right) \bm{E}_1 +
  \frac{c}{2k_{01}} \left[ \nabla^2 -
  (\hat{\bm{k}}_{01}\cdot\bm{\nabla})^2\right] \bm{E}_1 
  +
   \frac{\lambda ck_{01}}{2}\left( \frac{4}{3}\mathscr{E}_g +
  \alpha_1\mathscr{E}_2 \right)\bm{E}_1 = 0 , \nonumber \\ 
  \
\end{eqnarray}
and
\begin{eqnarray}
  i\left( \frac{\partial}{\partial t} +
  c\hat{\bm{k}}_{02}\cdot\nabla
  \right)\bm{E}_2 +
  \frac{c}{2k_{02}}\left[ \nabla^2 -
  (\hat{\bm{k}}_{02}\cdot\bm{\nabla})^2\right]\bm{E}_2 
   +
   \frac{\lambda ck_{02}}{2}\left( \frac{4}{3}\mathscr{E}_g +
  \alpha_2\mathscr{E}_1 \right)\bm{E}_2 = 0 ,\nonumber \\ 
  \
\end{eqnarray}
\label{eq:nlse}
\end{subequations}
were derived. They describe the propagation of two electromagnetic
pulses $\bm{E}_1$ and $\bm{E}_2$ on an incoherent radiation background. 
Here $\bm{k}_{0j}$ ($j = 1, 2$) is the unperturbed vacuum wave vector
(with a hat denoting the corresponding unit vector), $\lambda  
= 14\kappa$ or $8\kappa$ depending on the photon polarisation state,
while
\begin{eqnarray}
  \alpha_{1,2} = 2 -
   2\hat{\bm{k}}_{01}\cdot\hat{\bm{k}}_{02} -
   (\hat{\bm{k}}_{01,2}\cdot\hat{\bm{e}}_{02,1})^2 
   -
   [\hat{\bm{k}}_{01,2}\cdot(\hat{\bm{k}}_{02,1}\times\hat{\bm{e}}_{02,1})]^2
   , 
\end{eqnarray}
depends on the relative polarisation and propagation directions of the
two pulses in vacuum. Moreover, $\mathscr{E}_g$ and $\mathscr{E}_i =
\varepsilon_0\langle|E_i|^2\rangle$ is the energy density of the
radiation gas and 
the pulse $i$, respectively. Here the angular brackets denote the ensemble
average. 

We note that in the co-linearly propagating stationary case, Eq.\
(\ref{eq:wave2}) below yields $\mathscr{E}_g = 
2\lambda\mathscr{E}_0(\mathscr{E}_1 + \mathscr{E}_2)$, while Eqs.\
(\ref{eq:nlse}) may be written as 
\begin{subequations}
\begin{eqnarray}
  i\epsilon\frac{\partial E_1}{\partial x} +
  \frac{1}{2k_{01}}\nabla_{\perp}^2 E_1 
  +
   \lambda k_{01}\left[ \frac{4}{3}\lambda\mathscr{E}_0(\mathscr{E}_1
  + \mathscr{E}_2)  +
  \frac{1}{2}\alpha_1\mathscr{E}_2 \right]E_1 = 0 , 
\end{eqnarray}
and
\begin{eqnarray}
  i\frac{\partial E_2}{\partial x} +
  \frac{1}{2k_{02}}\nabla_{\perp}^2 E_2 
   +
   \lambda k_{02}\left[ \frac{4}{3}\lambda\mathscr{E}_0(\mathscr{E}_1
  + \mathscr{E}_2)  +
  \frac{1}{2}\alpha_2\mathscr{E}_1 \right] E_2 = 0 ,
\end{eqnarray}
\label{eq:nlse-stationary}
\end{subequations}
where $\epsilon = \hat{\bm{k}}_{01}\cdot\hat{\bm{k}}_{02}$, and we
have chosen the direction of propagation along the $x$-axis. Note that 
$\epsilon = \pm 1$ depending on whether the pulses are parallel or anti-parallel. For
parallel propagating beams, $\alpha_{1,2} = 0$, and the direct coupling
between the pulses vanishes. Still, because of the response of the
radiation background, the pulses are coupled, and Eqs.\
(\ref{eq:nlse-stationary}) exhibit spatial self-focusing
\cite{Kivshar-Agrawal}. Thus, we here generalise the two-dimensional
self-focusing results found perturbatively in Ref.\
\cite{Marklund-Brodin-Stenflo} and numerically in Ref.\
\cite{Shukla-Eliasson}. The results of the two-dimensional
self-focusing due to photon--photon scattering can also be understood
in the context of the 
modulational instability exhibited by the NLSE \cite{Shukla-Eliasson}.

The equations (\ref{eq:nlse}) are coupled to the acoustic wave
equation \cite{two-pulse}
\begin{eqnarray}
  && 
  \frac{\partial^2\mathscr{E}_g}{\partial t^2} -
  \frac{c^2}{3}\nabla^2\mathscr{E}_g =
  -\frac{2}{3}\lambda\mathscr{E}_0 \Bigg\{ 
  \left( 1 +
  \frac{\beta}{2}\sqrt{\frac{\mathscr{E}_2}{\mathscr{E}_1}}
  \right)\left( \frac{\partial^2}{\partial t^2} + c^2\nabla^2
  \right)\mathscr{E}_1  \nonumber  \\ &&
  + \left( 1 +
  \frac{\beta}{2}\sqrt{\frac{\mathscr{E}_1}{\mathscr{E}_2}}
  \right)\left( \frac{\partial^2}{\partial t^2} + c^2\nabla^2
  \right)\mathscr{E}_2  \nonumber \\ &&
  - \frac{\beta}{4}\sqrt{\frac{\mathscr{E}_2}{\mathscr{E}_1^3}} 
  \left[ \left( \frac{\partial\mathscr{E}_1}{\partial t}\right)^2 +
  c^2|\nabla\mathscr{E}_1|^2  \right]
  - \frac{\beta}{4}\sqrt{\frac{\mathscr{E}_1}{\mathscr{E}_2^3}} 
  \left[ \left( \frac{\partial\mathscr{E}_2}{\partial t}\right)^2 +
  c^2|\nabla\mathscr{E}_2|^2  \right] \nonumber \\ &&
%
  + \frac{\beta}{2\sqrt{\mathscr{E}_1\mathscr{E}_2}}
  \left[ \frac{\partial\mathscr{E}_1}{\partial
  t}\frac{\partial\mathscr{E}_2}{\partial t} +
  c^2(\bm{\nabla}\mathscr{E}_1)\cdot(\bm{\nabla}\mathscr{E}_2) \right] 
  \Bigg\} ,
\label{eq:wave}
\end{eqnarray}
for the radiation gas energy density $\mathscr{E}_g$,\footnote{Note
  that we have made the split $\mathscr{E}_g \rightarrow \mathscr{E}_0
  + \mathscr{E}_g$, with $\mathscr{E}_g \ll \mathscr{E}_0$, in
  accordance with Ref.\ \cite{two-pulse}, and transformed
  away the resulting phase shift term in Eqs.\ (\ref{eq:nlse}).}
where $\beta = \hat{\bm{e}}_1\cdot\hat{\bm{e}}_2 +
   (\hat{\bm{k}}_1\cdot\hat{\bm{k}}_2)\hat{\bm{e}}_1\cdot\hat{\bm{e}}_2 
   - 
   (\hat{\bm{k}}_1\cdot\hat{\bm{e}}_2)(\hat{\bm{k}}_2\cdot\hat{\bm{e}}_1)$.

\section{Kinetic description}

We assume that the polarisation of the pulses remains constant, and
define the Wigner functions $\varrho_j$ as the Fourier
transform of the spatial coherence function of $E_j$, $j = 1, 2$
\cite{Wigner,Moyal,Semikoz}   
\begin{eqnarray}
  \varrho_i(t,\bm{r},\bm{\kappa}) &=& \frac{1}{(2\pi)^3}\int\,d\bm{y}\,
  \exp(i\bm{\kappa}\cdot\bm{y}) 
  \langle E_i^*(t,\bm{r} +
  \bm{y}/2)E_i(t,\bm{r} - \bm{y}/2) \rangle ,
\label{eq:Wigner-transform}
\end{eqnarray}
where $\hbar\bm{\kappa}$ can be viewed as representing the momentum of
the individual photons. 
The Wigner function $\varrho_j$ has the property 
\begin{equation}
  \langle|E_j|^2\rangle =
  \int\,d\bm{\kappa}\,\varrho_j(t,\bm{r},\bm{\kappa}) .
\label{eq:normalisation}
\end{equation}

The transform (\ref{eq:Wigner-transform}) in conjunction with Eqs.\
(\ref{eq:nlse}) leads to the Wigner--Moyal equation
\begin{equation}
  \frac{\partial\varrho_j}{\partial t} +
  \left[c\hat{\bm{k}}_{0j} +
  \frac{c}{k_{0j}}\bm{\kappa} 
  - \frac{c}{k_{0j}}(\hat{\bm{k}}_{0j}\cdot\bm{\kappa})%
    \hat{\bm{k}}_{0j}\right]\cdot\bm{\nabla}\varrho_j +
  2\mathscr{U}_j\sin\left(  
  \frac{1}{2}\overleftarrow{\bm{\nabla}}\cdot%
    \overrightarrow{\bm{\nabla}}_{\kappa} \right)\varrho_j = 0 .
\label{eq:Wigner}
\end{equation}
where the potentials are defined according to
\begin{equation}
  \mathscr{U}_1 = \lambda ck_{01}\left( \frac{2}{3}\mathscr{E}_g +
    \frac{1}{2}\alpha_1\mathscr{E}_2 \right)~~\text{and}~~
  \mathscr{U}_2 = \lambda ck_{02}\left( \frac{2}{3}\mathscr{E}_g +
    \frac{1}{2}\alpha_2\mathscr{E}_1 \right)  
\end{equation}
and the intensities $\langle|E_i|^2\rangle$ are given by Eq.\
(\ref{eq:normalisation}). Moreover, the $\sin$-operator in Eq.\
(\ref{eq:Wigner}) is defined in terms of its Taylor expansion, the
arrows indicate the direction of operation, and
$\bm{\nabla}_{\kappa}$ denotes the derivative with respect to
$\bm{\kappa}$.  

We will now assume that the fields have perpendicular polarisations,
i.e. $\beta = 0$, and $\alpha_j = 0$ or $4$ if the beams are co- or
counter-propagating, respectively. Equation (\ref{eq:wave}) then simplifies
considerably, and takes the form
\begin{equation}
  \frac{\partial^2\mathscr{E}_g}{\partial t^2} -
  \frac{c^2}{3}\nabla^2\mathscr{E}_g =
  -\frac{2}{3}\lambda\mathscr{E}_0\left( \frac{\partial^2}{\partial
  t^2} + c^2\nabla^2 \right)(\mathscr{E}_1 + \mathscr{E}_2) ,
\label{eq:wave2}
\end{equation}

\section{Stability analysis}

In order to analyse the stability of the system (\ref{eq:Wigner}) and
(\ref{eq:wave2}), we linearise according the following scheme. 
Let $\varrho_j = \varrho_{0j}(\bm{\kappa}) +
\tilde{\varrho}_{j}(\bm{\kappa})\exp(i\bm{K}\cdot\bm{r} - i{\mno}t)$
(and the corresponding expression for $\mathscr{E}_j$), where
$\tilde{\varrho}_j \ll \varrho_{0j}$, 
and
$\mathscr{E}_g = \tilde{\mathscr{E}}_g\exp(i\bm{K}\cdot\bm{r} -
i{\mno}t)$. For co-propagating waves, Eqs.\ (\ref{eq:Wigner}) (in the
Vlasov limit) and
(\ref{eq:wave2}) give the dispersion relation
\begin{equation}
1 = W_1I_1^+ + W_2I_2^+ ,
\end{equation}
while in the case of counter-propagating pulses, we obtain
\begin{equation}
  \left( \tfrac{2}{9}\lambda\mathscr{E}_0\Delta\right)^2(1 - W_1I_1^+
  - W_2I_2^-) = \left(1 +
  \tfrac{4}{9}\lambda\mathscr{E}_0\Delta\right)W_1W_2I_1^+I_2^- 
\end{equation}
where $W_j \equiv (4/9)\lambda^2ck_{0j}\Delta$, $\Delta \equiv
({\mno}^2 + c^2K^2)/({\mno}^2 - c^2K^2/3)$, and
\begin{equation}
  I_j^{\pm} \equiv
  \int\,d\bm{\kappa}\,\frac{\bm{K}\cdot\bm{\nabla}_{\kappa}%
    \varrho_{0j}}{{\mno} \mp cK_x(1 - \kappa_x/k_{0j}) -
  (c/k_{0j})\bm{K}\cdot\bm{\kappa}} ,
\label{eq:integral}
\end{equation}
where $K_x = \hat{\bm{k}}_0\cdot\bm{K}$
For $\varrho_2 = 0$, we obtain a result similar to that of Ref.\
\cite{santorini} concerning Landau damping.

For mono-energetic beams, i.e., $\varrho_{0j}(\bm{\kappa}) =
\langle|E_{0j}|^2\rangle\delta(\bm{\kappa} - \bm{k}_{0j}) $, the
integral (\ref{eq:integral}) can be reduced, and we obtain the
well-known beam modulational instability. In the case of parallel
propagating beams, with $E_{01} = E_{02} \equiv E_0$, $k_{01} = k_{02}
\equiv k_0$, and
$\bm{\kappa}_{01} = \bm{\kappa}_{02} \equiv {k}_0\hat{\bm{x}}$, we obtain
the growth rate \cite{two-pulse}
\begin{equation}
 {\mng} \simeq \frac{1}{2}{cK_{\perp}}\sqrt{
 \frac{16}{3}\eta_0\eta_p\frac{2K_{x}^2
 + K_{\perp}^2}{-2K_{x}^2 + K_{\perp}^2} - \frac{K_{\perp}^2}{k_0^2}} ,
\end{equation}
where the dimensionless parameters are defined according to $\eta_0 =
\lambda\mathscr{E}_0$ and $\eta_p =
\lambda\mathscr{E}_p$. Furthermore, ${\mno}
= cK_{x} + i{\mng}$ and $\bm{K}_{\perp} = \bm{K} - K_x\hat{\bm{x}}$.

In general, however, the background distribution is not mono-energetic,
and due to the spectral width of $\varrho_0$ we will have Landau-like
damping manifested by the poles of the integral (\ref{eq:integral}). 
In the one-dimensional case, we can take the spectral width $\kappa_W$
into
account by considering two equal incoherent backgrounds with
$\varrho_0(\kappa) =
\langle|E_0|^2\rangle/(\sqrt{2\pi}\kappa_W)\exp[-\kappa^2/2\kappa_W^2]$,
so that $I^+ = 0$, and
\begin{eqnarray}
  I^- &=& \frac{\langle|E_0|^2\rangle k_0}{2c\kappa_W}\left\{ 1 +
  \frac{i\sqrt{\pi}k_0(\mno + c K)}{2\sqrt{2}c\kappa_W
  K}\,\text{erf}\left[ i\frac{k_0(\mno + c K)}{2\sqrt{2}c\kappa_W K}
  \right] \exp\left[ -\frac{k_0^2(\mno + c K)^2}{8c^2\kappa_W^2 K^2}
  \right] 
  \right\} \nonumber \\ &&
  + i\frac{\sqrt{\pi}\langle|E_0|^2\rangle k_0^2}{4c^2\kappa_W^3}
  \frac{(\mno + c K)}{\sqrt{2} K}\exp\left[
  -\frac{k_0^2(\mno + c K)^2}{8c^2\kappa_W^2 K^2} 
  \right] .
\end{eqnarray}
Here we clearly see that the
effect of the non-zero spectral width is to introduce a damping.

\section{Discussion and conclusion}

The existence of a modulational instability for mono-energetic
beams strongly suggests that the full nonlinear effect due to
photon--photon scattering of incoherent waves in a
radiation background should be taken into account, since any small
perturbation can grow to form a large amplitude structure, given
sufficient time. On the other hand, as the example with a
one-dimensional Gaussian distribution shows, we must also expect the
evolution to be damped by the resonant interaction between incoherent
modes and fluid modes. Thus, it is not unlikely that the full system
can account for very interesting structure formation, where the initial
growth is governed by a modulational type of instability, and that
this growth is stabilised by the Landau-like damping at a
later stage of the dynamical phase of the photon system. Thus, one can
therefore conjecture about the existence of three-dimensional stable
photon structures generated by the vacuum nonlinearities. In order to
investigate this conjecture, a numerical analysis of the full system
would be necessary. 

Situations where the effects presented in this paper may occur range from
earth-based laboratory systems, such as ultra-high intensity lasers
\cite{Mourou-Barty-Perry,Pukhov} 
and plasma accelerators \cite{Bulanov-Esirkepov-Tajima,Shen-etal},
to astrophysical scenarios, such as the early 
Universe \cite{Marklund-Brodin-Stenflo}, gamma ray bursts \cite{Piran}
and magnetars \cite{magnetar}. If the presents effects do occur, they could
also show up on small angular scales within high precision cosmology
measurements, such as the ones presented by WMAP
\cite{wmap,wmap2,wmap3}. A common feature for all these systems is
that they at some stage of their evolution lead to extreme radiation
energy densities, which is a  key feature for probing photon--photon
scattering. Thus, there are good possibilities that in
the near future one can put QED through new tests, both in laboratory
and astrophysical environments, where the latter is perhaps the most
fascinating, since it will connect very small scales to very
large scales as a fundamental theory test-bed. 

In the present paper we have used a system of nonlinear 
Wigner--Moyal equations. These equations are coupled through a
radiation fluid background, where the dynamics is determined by an
acoustic wave equation driven by the incoherent photons. Using these
equations, we have shown that incoherent electromagnetic pulses can
transfer energy between each other by means of the radiation fluid
background. Consistent with previous work, we have moreover shown that
the resulting system of equations is subject to a modulational
instability, and that a nonzero spectral width of the background
incoherent photons gives rise to a damping, much like the well-known
Landau-damping of electrostatic waves in plasmas. 
The implications of the results have been discussed, and it was conjectured
that stable three-dimensional electromagnetic
structures may form as a result of photon--photon scattering.

\end{document}